\documentclass[prb,
twocolumn,
superscriptaddress,showpacs,amsmath,amssymb]{revtex4}

\begin{document}

\title{On dimension of tetrads in effective gravity.}

\author{G.E.~Volovik}
\affiliation{Low Temperature Laboratory, Aalto University,  P.O. Box 15100, FI-00076 Aalto, Finland}
\affiliation{Landau Institute for Theoretical Physics, acad. Semyonov av., 1a, 142432,
Chernogolovka, Russia}

\date{\today}

\begin{abstract}
Two different sources of emergent gravity lead to the inverse square of length dimension of metric field, $[g_{\mu\nu}]=1/[l]^2$, as distinct from the conventional dimensionless metric, $[g_{\mu\nu}]=1$, for $c = 1$. In both scenarios all the physical quantities, which obey diffeomorphism invariance, such as the Newton constant, the scalar curvature, the cosmological constant, particle masses, fermionic and scalar bosonic fields, etc., are dimensionless.
\end{abstract}
\pacs{
}

\maketitle

\section{Introduction}:

There are several scenarios of emergent gravity. Gravity may emerge in the vicinity of the topologically stable Weyl point
\cite{Nielsen1976,Volovik1986,Froggatt1991,Horava2005,Volovik2003}; the analog of curved spacetime emerges in hydrodynamics with the so-called acoustic metric for the propagating sound waves\cite{Unruh1981};  etc. Here we consider two very different scenarios, which however have unusual common property: the tetrad fields in these theories have dimension of inverse length. As a result all the physical quantities which obey diffeomorphism invariance are dimensionless. This was first noticed by Diakonov \cite{Diakonov2011} and Vladimirov and Diakonov (VD)\cite{VladimirovDiakonov2012,VladimirovDiakonov2014} in the scenario, where tetrad fields emerge as bilinear combinations of the fermionic fields. Tetrads with  dimension of inverse length emerge also in the model of the superplastic vacuum \cite{NissinenVolovik2018,NissinenVolovik2019}.

\section{ Tetrads as bilinear combinations of fermion fields}:

In the theory by Vladimirov and Diakonov (VD) \cite{Diakonov2011,VladimirovDiakonov2012,VladimirovDiakonov2014} the tetrads are composite fields, which emerge as the bilinear combinations of the fermionic fields:
\begin{equation}
e^A_\mu=i \left<\psi^\dagger \gamma^A \nabla_\mu \psi + \nabla_\mu\psi^\dagger \gamma^A  \psi \right>
\,.
\label{bilinear}
\end{equation}
This construction is similar to what happens in the spin-triplet $p$-wave superfluids in the $^3$He-B phase \cite{Volovik1990}. In the VD scenario two separate Lorentz group of coordinate and spin rotations is spontaneously broken to the combined  Lorentz symmetry group, 
$L_{L}\times L_{S}\rightarrow L$. In the same manner in $^3$He-B the symmetries under three-dimensional rotations in orbital and spin spaces are broken to the symmetry group of combined rotations, $SO(3)_{L}\times SO(3)_{S}\rightarrow SO(3)_{J}$. 

Formation of tetrads spontaneously breaks both the symmetries under discrete coordinate transformations $P_{L}=({\bf r}\rightarrow -{\bf r})$ and $T_{L}=(t\rightarrow -t)$, and also the discrete symmetries in spin space, $P_{S}$ and $T_{S}$. 
The symmetry breaking scheme $P_{L}\times P_{S}\rightarrow P$ and $T_{L}\times T_{S}\rightarrow T$ leaves the combined parity $P$ and the combined time reversal symmetry $T$. 

The VD symmetry breaking mechanism can be important for the consideration of the Big Bang scenario, in which the gravitational tetrads change sign across the singularity, $e^A_\mu(\tau, {\bf x})=-e^A_\mu(-\tau, {\bf x})$ \cite{Turok2018,Turok2018b}. The singularity can be avoided by formation of the bubble with a vanishing determinant of the metric \cite{Klinkhamer2019,KlinkhamerLing2019},
which would correspond of the vacuum state with   unbroken symmetry, i.e.  with zero tetrad field, $e^A_\mu=0$.
On the other hand, the Big Bang can be considered as a symmetry breaking phase transition $L_{L}\times L_{S}\rightarrow L$, at which the symmetry between the spacetime with $e>0$ and anti-spacetime  with $e<0$ is spontaneously broken, where $e$ is the tetrad determinant.

Correspondingly, in superfluid $^3$He the formation of the $p$-wave order parameter spontaneously breaks the symmetry under coordinate transformation ${\bf r}\rightarrow -{\bf r}$. The VD scenario has also the connection to the chiral $^3$He-A phase: in both systems the topologically protected Weyl fermions emerge, which move in the effective tetrad field \cite{Volovik2003}. 

According to Eq.(\ref{bilinear}), the frame field $e_\mu^A$ transforms as a derivative and thus has the dimension of inverse length, 
$[e_\mu^A]=1/[l]$ (it is assumed that $\psi$ is scalar under diffeomorphisms) \cite{Diakonov2011,VladimirovDiakonov2012}. The dimension of the metric is 
$[g_{\mu\nu}]=1/[l]^2$. 
For Weyl or massless Dirac fermions one has the conventional action:
\begin{equation}
S= \int d^4x |e| e^{A\mu}\left(\psi^\dagger \gamma^A \nabla_\mu \psi + {\rm H.c.} \right)
\,.
\label{DiracZeroMass}
\end{equation}
The action (\ref{DiracZeroMass}), when expressed in terms of the Vladimirov-Diakonov tetrads, is dimensionless, since $[e]=[l]^{-4}$, $[e^{A\mu}] =[l]$ and $[\psi]=1$. This suggests that the VD dimension of tetrads is natural, and  the tetrads emerging for example in the chiral Weyl superfluid  $^3$He-A  and in Weyl semimetals may also have the  dimension $1/[l]$.

\section{ Elasticity tetrads of superplastic vacuum}:

The elasticity tetrads describe elasticity theory \cite{DzyalVol1980,VolovikDotsenko1979, AndreevKagan84,NissinenVolovik2018,NissinenVolovik2019}. 
In conventional crystals they are gradients of the three $U(1)$ phase fields $X^A$, $A=1,2,3$,
\begin{equation}
e^{~A}_\mu(x)= \partial_\mu X^A(x)\,.
\label{reciprocal}
\end{equation}
The surfaces of constant phases, $X^A(x)=2\pi n^A$, describe the system of the deformed crystallographic planes.
Being the derivatives, elasticity tetrads have also canonical dimensions of  inverse length. This allows us to extend the application of the topological anomalies. For example, the Chern-Simons term describing the 3+1 intrinsic quantum Hall effect becomes dimensionless. As a result, the prefactor of this term is given by the integer momentum-space topological invariants in the same manner as in the case  of 2+1 dimension. In principle, the elasticity tetrads can be used as the gravitational tetrads for the construction of gravity in the model of the 3+1 vacuum as a plastic (malleable) fermionic crystalline medium with $A=0,1,2,3$ \cite{KlinkhamerVolovik2019,Zubkov2019}. 

\section{ Dimensionless world}:
 
In this superplastic vacuum based on elasticity tetrads, all the physical quantities become dimensionless \cite{NissinenVolovik2019}. The reason for that is that  in the superplastic vacuum, which can be arbitrarily deformed, the equilibrium size of the elementary cell is absent, and thus the microscopic length scale (such as Planck scale)  is absent. That is why all the distances are measured in terms of the integer positions of the nodes in the crystal. As a result, the gravitational constant $K$ (inverse Newton constant), the scalar curvature $R$, the cosmological constant $\Lambda$, and particle masses $M$ become dimensionless \cite{NissinenVolovik2019}.

The same is valid for gravity in terms of the Vladimirov-Diakonov tetrads,
where as they wrote "all world scalars are dimensionless, be it the scalar
curvature $R$, the interval $ds$, the fermion field $\psi$, or any
diffeomorphism-invariant action term" \cite{VladimirovDiakonov2012}. 
The particle masses $M$ (or mass matrices)  are also dimensionless. When the conventional mass term in the Dirac action
\begin{equation}
S= \int d^4x |e| M\psi^\dagger \psi 
\,,
\label{MassTerm}
\end{equation}
is expressed  in terms of the VD tetrads,  then from $[e]=[l]^{-4}$ and $[\psi]=1$ one obtains $[M]=1$.

Correspondingly, for bosonic scalar field
\begin{equation}
S= \int d^4x\sqrt{-g} \left( g^{\mu\nu} \nabla_\mu \Phi  \nabla_\nu \Phi + M^2 \Phi^2 \right)\,,
\label{Bosonic}
\end{equation}
one has $[\sqrt{-g}]=[l]^{-4}$, $[g^{\mu\nu}]=[l]^2$, $[\Phi]=1$ and $[M]=1$.
In particular,  the action for the Bose condensate phase $\Phi$  in superfluid $^4$He contains the effective acoustic metric with
 $\sqrt{-g}=(n/mc^2)^2$, where $n$ is particle density, $m$ is the mass of $^4$He atom, and $c$ is the speed of sound \cite{Volovik2003}.
This gives $[\sqrt{-g}]=[l]^{-3}[t]^{-1}$. The same is for the effective metric for the propagating Goldstone modes of the coherent spin precession \cite{NissinenVolovik2017}, and actually for any Goldstone mode with linear spectrum.
Note that when $c$ and $\hbar$ are incorporated into the tetrad fields and thus to the metric,
these quantities do not enter explicitly into the diffeomorphism-invariant action \cite{Volovik2009}.

\section{ Conclusion}:

In both scenarios of emergent gravity, the dimensionless physics is supported by the invariance under diffeomorphisms. In the VD  theory this invariance is assumed as fundamental.  In the superplastic vacuum, the diffeomorphism invariance corresponds to the proposed invariance under deformations of the 4D crystal. All this suggests that the dimensionless physics can be the natural consequence of the diffeomorphism invariance, and thus can be the property of the gravity, which we have in our quantum vacuum.

Note the difference with the conventional expression of the physical parameters in terms of the Planck units, where the Newton constant 
$G=1$, and all the physical quantities also become dimensionless. In this approach the masses of particles are expressed in terms of the Planck energy, which is assumed to be the fundamental constant. However, in principle the Planck energy or the Newton constant may depend on the trans-Planckian physics, and thus can (and should) be space and coordinate dependent (see, e.g. the time-dependence of $G$ in the so-called $q$-theory, when $G$ approaches its asymptotic value in the Minkowski vacuum \cite{KlinkhamerVolovik2008}). On the contrary, in the Vladimirov-Diakonov approach the ``fundamental constants`` do not exist, and only dimensionless ratios  and the topological quantum numbers make sense. Then, instead of the fundamental constants,  the most stable physical quantities should be used. The masses of particles or the Newton and cosmological ``constants`` can be expressed for example in terms of the Bohr radius \cite{VladimirovDiakonov2012}. 
Note that in the modified gravity theories, such as the scalar-tensor and $f(R)$ theories 
(see e.g. \cite{Starobinsky1980}), the effective Newton ``constant`` $G$ can be space-time dependent, and thus is not fundamental.
 
The dimensionless physics emerging in the frame of the Vladimirov-Diakonov dimensionful tetrads leads in particular to the new topological terms in action, since some of the dimensionless parameters appear  to be the integer valued quantum numbers, which chacterize the topology of the quantum vacuum. This can be seen on example of the 3+1 dimensional quantum Hall effect in topological insulators
\cite{NissinenVolovik2018,NissinenVolovik2019,Vishwanath2019}. When the Chern-Simons action is written in terms of  the elasticity tetrads with $[e_\mu^A]=1/[l]$, its prefactor becomes dimensionless and universal, being expressed in terms of  integer-valued momentum-space invariant. 

The relativistic example of such phenomenon is the chiral anomaly in terms of the torsion fields suggested by Nieh and Yan \cite{NiehYan1982a,NiehYan1982b,Nieh2007}. For the conventional torsion and curvature in terms of the conventional dimensionless tetrads, the gravitational Nieh-Yan anomaly equation for the non-conservation of the axial current
\begin{equation}
\partial_\mu   j_5^\mu =  \lambda^2
\left(\mathcal{T}^A \wedge \mathcal{T}_A - e^A \wedge e^B\wedge R_{AB} \right) \,, 
\label{eq:NYterm}
\end{equation}
contains the nonuniversal prefactor -- the ultraviolet cut-off parameter $\lambda$ with dimension of inverse length, $[\lambda]=1/[l]$. Because of such prefactor,  the Nie-Yan contribution to the anomaly is still contentious and subtle (see recent literature 
\cite{Khaidukov2018,Nissinen2019,NissinenVolovik2019b,Stone2019,Ojanen2019,ZeMin2020}). This is  because the nonuniversal parameter may depend on the spacetime coordinates, which explicitly violates the topology.
However, in terms of Vladimirov-Diakonov tetrads, the dimension of torsion becomes $[\mathcal{T}_A]=1/[l]^2$, and as a result the prefactor 
$\lambda$ becomes dimensionless, $[\lambda]=1$. The latter suggests that the prefactor is universal, and thus properly  reflects the topology of the quantum vacuum. 

 {\bf Acknowledgements}. This work has been supported by the European Research Council (ERC) under the European Union's Horizon 2020 research and innovation programme (Grant Agreement No. 694248).

\end{document}